\documentstyle[prb,aps]{revtex}
\begin{document}
\preprint{Bicocca 3/FT,  Jan. 1999}
\draft
\title
 {Critical specific heats of the $N$-vector spin models \\
 on the sc and the bcc lattices }
\author{P. Butera\cite{pb} and M. Comi\cite{mc}}
\address
{Istituto Nazionale di Fisica Nucleare\\
Dipartimento di Fisica, Universit\`a di Milano\\
16 Via Celoria, 20133 Milano, Italy}
\maketitle
\scriptsize
\begin{abstract}

 We have   computed  through   order
$\beta^{21}$  the high temperature expansions for the nearest 
neighbor spin correlation function   $G(N,\beta)$
of the classical $N$-vector model, with  general $N$, on
 the simple-cubic  and on the body-centered-cubic lattices.  
 For this model, also known in quantum 
field theory as the lattice $O(N)$ nonlinear sigma model, 
we have presented in previous papers extended expansions of the 
susceptibility, of its second field derivative and of the second 
moment of the correlation function.
 Here we study the internal specific energy and
  the specific heat $C(N,\beta)$,   
 obtaining  new estimates of the critical parameters 
and therefore a more accurate direct test of the hyperscaling relation 
$d \nu(N)=2 - \alpha(N)$  on a range of  values of the spin 
dimensionality $N$, including
 $N=0$ [the self-avoiding walk model], $N=1$ [the Ising spin 1/2 model], 
 $N=2$ [the XY model],  $N=3$ [the classical Heisenberg model].
By the newly  extended series, 
 we also compute the universal combination of critical amplitudes
  usually denoted by $R^+_{\xi}(N)$,  
 in fair agreement with renormalization group estimates.
\end{abstract}
\normalsize
\widetext

\section{Introduction}
 We continue in this note the analysis of 
 recently extended \cite{bc2d,bc3d,bcgstar} high temperature (HT) 
expansions  for the $N$-vector model\cite{st68} with general 
 spin dimensionality $N$.
Our computation is concerned with the $d$-dimensional bipartite lattices, 
namely  the simple-cubic (sc) lattice,  the body-centered-cubic (bcc) 
lattice and their $d-$dimensional generalizations.

In  previous papers we  have tabulated:  
 i) the HT series for the zero field susceptibility $\chi(N,\beta)$
and for the second  moment of the correlation function $\mu_2(N,\beta)$
 through order $\beta^{21}$, 
 ii) the HT series for the second field derivative
of the susceptibility  $\chi_4(N,\beta)$ through order  $\beta^{17}$,
 and  have analysed their critical behavior
in the $d=2$ case\cite{bc2d} and in the $d=3$ case\cite{bc3d,bcgstar}.
 We have performed the computation using the (vertex-renormalized) linked 
cluster expansion  method\cite{w74mck83} and  have
 produced  tables of series  
coefficients written as explicit functions of 
the spin dimensionality $N$   with an 
extension independent of the structure and 
dimensionality of the  lattice.
 More details on the derivation of the series, and on 
 the checks of validity of  our results
 can be found in our 
previous papers\cite{bc2d,bc3d,bcgstar}. 

 In this paper we examine the series expansions of the 
 nearest neighbor correlation function $G(N,\beta)$  
 through order $\beta^{21}$,  in order 
to update, on a range of  values of the spin 
dimensionality $N$, the direct estimates of  the  
parameters describing the behavior of the specific heat $C(\beta,N)$ 
on the HT side of the critical point $\beta_c(N)$. 
 We also update  direct tests of the hyperscaling relation 
$d \nu(N)=2 - \alpha(N)$    and  estimate a related universal combination
 of critical amplitudes introduced by Stauffer, Ferer and Wortis\cite{stau}
 and later denoted by $R^+_{\xi}(N)$\cite{hohe,ber,aha,zinn}.
Here $\alpha(N)$ is the critical exponent of the specific 
heat    and  $\nu(N)$  is the critical exponent 
 of the correlation length $\xi(N,\beta)$.
 Estimates  of  $\alpha(N)$ are also  obtained   by 
studying the behavior of the extended  series for the 
susceptibility $\chi(\beta,N)$ and for 
the second moment of the correlation function 
$\mu_2(\beta,N)$ at the symmetrically placed 
anti-ferromagnetic singular point $\beta^{af}_c(N)= -\beta_c(N)$\cite{fich}.

 In order to put our work into a proper perspective, it 
is convenient to list  the  HT expansions of $G(N,\beta)$ 
for the sc,  the bcc and the face-centered-cubic (fcc) lattices, which
were published before our extension.
As well known, for $N=0$,  the  $N$-vector model reduces\cite{gen,hug} to
 the self-avoiding walk (SAW) model, and the expansion
  of the correlation function $G(0,\beta)$, simply related to the 
enumeration of the self-avoiding rings (or polygons)
  had already been computed  in Ref.\cite{smwm} up to
 order $\beta^{19}$ for the sc lattice,  up to 
order $\beta^{15}$ for the bcc
lattice and up to $\beta^{13}$ for the fcc lattice. 
In the   $N=1$ case, which corresponds to 
 the spin 1/2 Ising model, an expansion of $G(1,\beta)$  
for the sc lattice 
 has been obtained  a few years ago by Enting and Guttmann 
\cite{ge} up  to order $\beta^{21}$ using
 finite lattice methods. More recently, within the same approach, this 
computation has been pushed to  order $\beta^{23}$ in 
Ref.\cite{bcgs} and then to 
 order $\beta^{25}$ in Ref.\cite{ge2}. 
Also an approximate determination of the 
coefficient of $\beta^{27}$ is reported in the last Reference.
  An expansion through  order $\beta^{15}$ for the bcc lattice,
and  one
 for the fcc lattice up to 
order $\beta^{12}$ have been  tabulated in Ref.\cite{shmh}. 
For  $N=2$ (the $XY$ model) the available series\cite{fmw}
 for  the bcc 
lattice reached  the order $\beta^{11}$. 
In the  $N=3$ case (the classical Heisenberg 
model), the series for the bcc lattice,   
 known only up to  order $\beta^{9}$, is reported  in Ref.\cite{rifi}.
 
Finally, let us cite  
 an expansion of $G(N,\beta)$,  valid for general $N$ and
 for all loosely packed lattices,  
 tabulated (with some misprints)
 in Ref.\cite{st68} up to order  $\beta^{9}$, which
 has been later extended to models with 
general anisotropic pair interaction in Ref.\cite{gerb}.
 The expansion of $G(N,\beta)$ has been 
recently pushed\cite{pisa}  to order  $\beta^{15}$ in the case of the sc 
lattice, 
 but no comparable effort has   been devoted
to the  bcc lattice.
In Ref.\cite{engl}, an expansion to order  $\beta^{11}$, 
valid for general $N$,
  had been tabulated 
 for the fcc lattice.

We should finally call the readers' attention to the valuable reviews 
 in Refs.\cite{aha,zinn,kumar} and to the  accurate
 recalculation,  within the Renormalization Group (RG) approach, 
of the universal critical 
parameters of the $N-$vector model performed by 
Guida and Zinn-Justin\cite{gz2}. This work is 
 based  on the recently extended field theoretic expansions 
of Ref.\cite{mur}   and  is also  accompanied by an extensive 
 review of the  available numerical and experimental data.

 The paper is organized as follows. In Sec.
II we set our notation and
 define the quantities  we shall study.
 
 In Sec. III we  discuss briefly the numerical
 tools  used for our estimates and present the results of 
 our analysis of the series. 
 These results are compared with experimental data,
 with  earlier work on shorter HT series,  with 
 measures performed in   stochastic simulations and
 with  RG estimates, obtained either
 by the fixed dimension (FD) perturbative  technique
\cite{zinn,gz2,mur,zib,zib1,brez,itdr,babe}
 or by the Fisher-Wilson\cite{epsi}
$\epsilon$-expansion approach\cite{zinn,gz2,brez,itdr,babe,epsi1,epv}.

 Our conclusions are briefly summarized in Sec. IV.

 The HT series expansion  coefficients 
 of the nearest neighbor correlation function 
 $G(N,\beta)$ expressed in  closed form
as functions of the spin dimensionality $N$, 
for the sc and the bcc lattices, 
 have been tabulated in the appendices
 in order to make each step of our work completely   reproducible.
For convenience of the reader, we also have  explicitly evaluated the 
 series coefficients for  $N=0$ (the  SAW model),
$N=1$  (the Ising spin 1/2 model),
 $N=2$ (the XY model) and $N=3$ (the classical Heisenberg model).

\section{ Definitions and notation}

We study  the $N$-vector model  with Hamiltonian:  

\begin{equation}
H \{ v \} = -{\frac {1} {2}} \sum_{\langle \vec x,{\vec x}' \rangle }
v(\vec x) \cdot v({\vec x}').
\label{hamilt} \end{equation}
where the variable $v(\vec x)$ represents a  $N$-component classical spin
of unit length at the lattice site  with position vector $\vec x$,
and the sum extends to all nearest neighbor pairs of  sites.

The basic observables are the spin correlation functions. Here we shall be 
interested  in   
the connected correlation functions $\langle v(0) \cdot v(\vec x) \rangle_c$
 between the spin at
the origin and the spin at the site $\vec x$.
In particular, the nearest neighbor spin 
correlation function is defined by

\begin{equation}
G^{\#}(N,\beta) =  \langle v(0) \cdot v(\vec \delta) \rangle_c = 
  \sum_{r=0}^\infty a^{\#}_{r}(N) \beta^{r}. \label{corr} \end{equation}
where $\vec \delta$ is a nearest neighbor lattice vector and  
$\#$ stands for either sc or bcc, as appropriate.

Due to the bipartite structure of the sc and the bcc lattices,
  the connected correlations 
 $\langle v(0) \cdot v(\vec x) \rangle_c$
 are    functions of $\beta$ with the same parity as the 
lattice distance between the 
spins and hence alternate expansion coefficients  vanish 
identically: in particular in our expansions 
 of $G^{\#}(N,\beta) $ to order $\beta^{21}$
  only eleven  coefficients  are nonvanishing. 
 This is the reason why most analyses in the 
literature have  focused on  
series for the non-bipartite fcc lattice 
which have no such symmetry.  

The specific internal energy is defined by
\begin{equation}
U^{\#}(N,\beta)= -\frac{q} {2} G^{\#}(N,\beta) 
\end{equation}
 where   $q$ is the lattice coordination number. 

If we denote the reduced inverse temperature by 
$\tau^{\#}(N) =1 - \beta/\beta^{\#}_c(N) $,
 then $U^{\#}(N,\beta)$ is expected to behave  as\cite{aha}
\begin{equation}
 U^{\#}(N,\beta)
\simeq U^{\#}_{reg}(N,\beta)+ A^{\#}_{U}(N)(\tau^{\#}(N))^{1-\alpha(N)}
\Big(1+ a^{\#}_{U}(N)
(\tau^{\#}(N))^{\theta(N)} +...\Big)
\label{confu}\end{equation}
when $ \tau^{\#}(N) \downarrow 0$.

As  customary, in writing the asymptotic form  eq. (\ref{confu}),  
 we have explicitly indicated 
 the presence of the non-singular background 
$U^{\#}_{reg}(N,\beta)$, 
 because  the critical  singularities of the specific energy are 
 known to be generally very weak. 
 Here $A^{\#}_{U}(N)$ denotes the critical amplitude of 
 the specific energy,
  $a^{\#}_{U}(N)$ is the amplitude of the leading singular 
correction\cite{wegner} 
    to scaling, $\theta(N)$ is the exponent of this
correction  also called confluent singularity exponent. 
The ellipses represent higher order singular or analytic correction terms.
 Unlike the critical exponent $\alpha(N)$, which is universal,  
 the critical
amplitudes $A^{\#}_{U}(N), a^{\#}_{U}(N),$ etc. are expected to
 depend  on the parameters of the Hamiltonian and on the lattice structure,
i.e. they are non-universal.
 Similar considerations  apply
 to the other  thermodynamic quantities listed below,
 which have  different critical
 exponents and different critical amplitudes, but
the  same leading confluent exponent $\theta(N)$.
  It is known that  $\theta(N) \simeq 0.5$
for small values of  $N$\cite{zinn}.
  Having clearly indicated which quantities
are universal,  we shall often
 drop  the generic superscript   $\#$ (or its determination) 
in order to avoid overburdening the notation.
 Notice also that, since there is no chance
of confusion, we have  generally omitted the superscript
$+$ usually adopted in the literature
  for the amplitudes which characterize the high
temperature side of the critical point.

The specific heat per site, at fixed magnetic field $H$,   is defined as
 the temperature derivative of the specific internal energy

\begin{equation}
C_{H}(N,\beta) = \frac{d} {dT} U(N,\beta) = \frac{q} {2} \beta^2 
\frac{d} {d\beta} G(N,\beta) 
 \label{csp} \end{equation}
where  $T$ is the temperature.
As $\tau(N)  \downarrow 0$, the critical behavior of $C_H(N,\beta)$ 
is described by 

\begin{equation}
 C_{H}(N,\beta)
\simeq C^{reg}_{H}(N,\beta)+ A_{C}(N)(\tau(N))^{-\alpha(N)}\Big(1+ a_{C}(N)
(\tau(N))^{\theta(N)} +....\Big)
\label{confc}\end{equation}
with $A_{C}(N)= (1-\alpha(N))\beta_c(N)A_U(N)$ and 
$a_C(N)=(1+ \frac {\theta(N)}{1-\alpha(N)})a_U(N)$.
 Notice that our definition of $a_C(N)$ 
conforms to general usage, but differs
 by a factor $\alpha(N)$ from eq. (1.4) of Ref.\cite{babe}. 

We have also examined the  susceptibility 

\begin{equation}
\chi(N,\beta) = \sum_{\vec x} \langle v(0) \cdot v(\vec x) \rangle_c,
 \label{chi} \end{equation}
   the second  moment of the correlation function  
\begin{equation}
 \mu_{2}(N,\beta)=\sum_{\vec x} \vec x^2 \langle v(0) \cdot v(\vec x)
\rangle_c,
\end{equation}
and  the second-moment 
 correlation length $\xi$ defined\cite{fibu}, 
in terms of $\chi$ and $\mu_{2}$, 
 by
\begin{equation}
 \xi^{2}(N,\beta)= \frac  {\mu_{2}(N,\beta)} {6\chi(N,\beta) }.
\end{equation}
    
The susceptibility $\chi(N,\beta)$ is expected to behave as 

\begin{equation}
 \chi(N,\beta)
\simeq A_{\chi}(N)(\tau(N))^{-\gamma(N)}\Big(1+ a_{\chi}(N)
(\tau(N))^{\theta(N)} +...\Big)
\label{confchi}\end{equation}                                         
as $\tau(N)  \downarrow 0$. In the case of bipartite lattices
 $\chi(N,\beta) $  has also an antiferromagnetic singularity at 
$\beta^{AF}_c(N)= -\beta_c(N)$, and, in terms of the reduced variable  
$\tilde \tau(N) =1 - \beta/\beta^{AF}_c(N) $,    we should observe 
the energy-like behavior

\begin{equation}
 \chi(N,\beta)
\simeq \chi_{reg}(N,\beta) + B_{\chi}(N)(\tilde 
\tau(N))^{1-\alpha(N)} + ...
\label{chiaf}\end{equation}
as $\tilde \tau(N)  \downarrow 0$.

The  second  moment of the correlation function is 
expected to behave as 

\begin{equation}
 \mu_{2}(N,\beta)
\simeq A_{\mu}(N)(\tau(N))^{-\gamma(N)- 2\nu(N)}\Big(1+ a_{\mu}(N)
(\tau(N))^{\theta(N)} +...\Big)
\label{confmu2}\end{equation}                                         
as $\tau(N)  \downarrow 0$. At the antiferromagnetic singularity,  
 the behavior is completely similar to that of the susceptibility
\begin{equation}
  \mu_{2}(N,\beta)
\simeq \mu_{2}^{reg}(N,\beta) + B_{\mu}(N)(\tilde 
\tau(N))^{1-\alpha(N)} + ...
\label{muaf}\end{equation}
as $\tilde \tau(N)  \downarrow 0$.

For the correlation length we have
\begin{equation}
 \xi(N,\beta)
\simeq A_{\xi}(N)(\tau(N))^{-\nu(N)}\Big(1+ a_{\xi}(N)
(\tau(N))^{\theta(N)} +...\Big)
\label{confxi}\end{equation}                                         
 as $ \tau(N)  \downarrow 0$, and also in this case 
 we expect the energy-like behavior
 \begin{equation}
  \xi(N,\beta)
\simeq \xi^{reg}(N,\beta) + B_{\xi}(N)(\tilde 
\tau(N))^{1-\alpha(N)} + ...
\label{xiaf}\end{equation}   
as $\tilde \tau(N)  \downarrow 0$.

 The validity of the hyperscaling relation 

\begin{equation}
d\nu(N) = 2 -\alpha(N)
\label{gunt}\end{equation}
 first derived by Gunton and Buckingham\cite{gunt} as an inequality
  (with the  = sign replaced by $\geq$),
translates into the universality 
of the amplitude combination\cite{stau} 

\begin{equation}
 R^+_{\xi}(N)  
\equiv \Big(g\alpha(N) A_C(N)\Big)^{1/d} A_{\xi}(N)
\label{egunt}\end{equation}
where $g$ is a geometric factor defined by 
$g=a^d/v_0$, with $v_0$ the volume per lattice site 
and $a$ the lattice spacing. For the sc lattice one has
$g=1$, while for the bcc lattice $g=3\sqrt3/4$. 

 Finally, it is useful to recall\cite{gerber} that,
 in the large $N$ limit, 
$ R^+_{\xi}(N)\approx \big(\frac {N} {4\pi}\big)^{1/3}$ 
and that Bervillier and 
Godr\`eche\cite{bg} proposed a simple approximate extension 
of this relationship
  to small nonzero values of  $N$ in the form 
$ R^+_{\xi}(N)\approx \nu(N)\big(\frac {N} {4\pi}\big)^{1/3}$.

\section{ Comments on the analysis of the series} 
\subsection{Estimates of the specific heat exponents}
 The main difficulty in computing 
the specific heat exponents  
  is that $\alpha(N)$ is  small for $N \leq 1$ and it
 becomes negative for $N \geq 2$. Therefore the specific heat 
 is very weakly divergent for $N \leq 1$, 
whereas it has only a finite cusp for $N \geq 2$. The simplest 
Pad\'e approximant (PA) techniques for estimating the critical parameters
 are thus expected  to be   inefficient in the former cases and completely 
inadequate in the latter. Moreover, 
 it is not particularly   helpful  to differentiate the present 
 specific heat series with respect to 
$\beta$ in order to sharpen the singularity, 
  because the extrapolations  become more sensitive to 
non-asymptotic or  confluent singularity effects. 
In principle, the inhomogeneous differential 
approximants (DA) (thoroughly described in Refs.\cite{gutasy})
 should perform much better than the PA's 
 since they are able to detect even   weak singularities and 
might allow,  to some extent, 
 for the confluent corrections to scaling.
 However, even after our extension of the HT  series, the nonzero expansion 
coefficients  are not    sufficiently many that these 
 numerical tools  can be used effectively. 
  In order to improve 
the  precision  of our estimates, 
we have mainly used simple first order DA's  
and have biased  them 
with the critical temperatures reliably known from our previous 
study of the  strongly divergent susceptibility 
series\cite{bc3d} or from other sources\cite{blh,nr90}. 
  In the particular case of the 
 sc spin 1/2 Ising model, we have 
 taken advantage in our analysis also of the two additional 
series coefficients  provided by Ref.\cite{ge}. 

  An accurate measure of the scaling correction 
 amplitudes of the specific heat presently seems   
beyond  reach, although 
their qualitative behavior as functions 
of $N$ is clear and completely analogous 
to that of $a_{\chi}(N)$ and of $a_{\xi}(N)$. 
 More precisely, $a^{\#}_C(N)$ is small and negative for $N<2$, 
 while it is positive and increasing for $N>2$. 
 Let us recall that, for small values of $N$,
 RG computations\cite{aha,zinn} indicate 
that  the universal ratios $a_C(N)/a_{\chi}(N)$ and 
$a_C(N)/a_{\xi}(N)$ are of the order of the unity. On the other hand,
 our HT analysis 
of $\chi(\beta,N)$ and $\xi(\beta,N)$ 
suggested that $a_{\chi}(N)$ and $a_{\xi}(N)$ are small (negative for 
 $N<2$ and positive otherwise),   
 therefore it is reasonable to  neglect the corrections to scaling 
at the present level of accuracy in the specific heat series analysis. 
 We also recall  that it was convincingly inferred in Ref.\cite{liuf}
 that $a_C$ is negative in the sc, bcc and fcc spin 1/2 Ising models 
 and, in the sc case, it was   suggested  
in Refs.\cite{ge,bcgs} that $a_C$ is very small. 

Our  direct  estimates of $\alpha(N)$ 
from  the specific heat series for the  sc and the bcc lattices
   have been reported  in Table 1.
 We have also included in this Table  the values of  $\alpha(N)$ 
 obtained by studying the energy-like  
behavior of the susceptibility eq.(\ref{chiaf}) at 
the antiferromagnetic singularity.   
 The study of the second correlation moment eq.(\ref{muaf}) 
 does not produce results of comparable quality.
 In this computation,  we have found most convenient to analyse 
 the derivative of $\chi$  by second order DA's 
biased   with 
the singularities at  $\beta_c(N)$ and  $\beta^{af}_c(N)$. 
 Although  the expansion 
of $\chi$  is effectively longer 
than that of the specific heat, it is not easier 
to measure  accurately the exponent of the very weak antiferromagnetic
 singularity. 
 Therefore the estimates of $\alpha(N)$ so obtained are 
 consistent with, but not 
more accurate than the others. In particular, we agree with the
 earlier estimates 
$\alpha(1)=0.105(7)$ and $\alpha(1)=0.11(2)$ 
obtained by studies of the susceptibility for the Ising model 
on the bcc lattice in Refs.\cite{fich,nr90}.

 In  recent studies of the $N=1$ case\cite{bcgs,ge2},  
it has been suggested that 
 the behavior of the specific heat series coefficients 
as functions of their order is 
sufficiently smooth that  the traditional 
(biased) ratio techniques 
 can be practically as accurate as the DA procedures. This  remains 
true only for not too large values  of $N$, since an 
asymptotic regime seems to set in  later for larger $N$. 
 Moreover, for $N> 4 $, the ratio
 sequences show an increasing curvature indicating that 
 the confluent corrections 
to scaling cannot be neglected anymore and therefore longer 
series are needed for a  reliable analysis.

We have used the simplest ratio formulas, since the more elaborate variants
 proposed in Ref.\cite{zinn79} do not presently make much difference. 
 If we set $C_H(N,\beta)= \sum_{n=1}^\infty c_{n}(N) \beta^{2n}$,  
 and allow for  the dominant 
 corrections to scaling with exponent $\theta(N)$, 
the ratio of the successive coefficients 
 of the specific heat  expansion in powers of $\beta^2$ 
is expected to behave as 
\begin{equation}
 r_n= \frac{c_n}{c_{n+1}}= \beta_c^2 \big(1+ \frac{1-\alpha} {n} + 
\frac{b}{ n^{1+\theta}} +O(\frac{1}{n^2})\big)
\end{equation}
Therefore $\alpha$ can be estimated from the sequence 

\begin{equation}
\alpha_n= 1- ( \frac{r_n}{\beta_c^2} -1 )n= \alpha + \frac{b}{ n^{\theta}}
+O(\frac{1}{n}) 
\end{equation}  

The extrapolation of these estimators
 to $n \rightarrow \infty$ is 
the main difficulty with this procedure.
 For $N\leq4$,  the estimators, when plotted 
versus $1/n$, show only a small
 curvature.  Therefore 
we have neglected the scaling correction $b/ n^{\theta}$ and have
  simply taken the linear extrapolant
 $n\alpha_n - (n-1)\alpha_{n-1}$ of the last two estimators
 as  our final estimate of $\alpha(N)$.
 We have then assigned very conservative 
uncertainties to these results 
(also allowing  for the errors in $\beta_c(N)$) 
and, for $N > 2$, 
we have indicated by asymmetric errors
the effects of some  curvature 
in the estimator plots.

In Table I, we have also included  the results of a few recent 
direct studies of the specific heat by 
stochastic  methods. These studies are subject to difficulties 
analogous to those met in HT analyses. As a consequence,  
 for instance, the MonteCarlo (MC) determination\cite{def} 
of $\alpha(0)$ on the sc lattice is 
approximately three standard deviations 
away from the other quoted values. (We have 
 quoted the sum of the systematic and the 
statistical errors separately reported  in Ref.\cite{def}.) 

Also  the value  of 
$\alpha(1)$ emerging  from a most accurate (see Ref.\cite{blh})
  MC study of the sc lattice Ising model 
 performed by a dedicated processor,  shows a considerable uncertainty. 
The central value, but not the error,  is somewhat improved 
($\alpha(1)=0.113 \pm 0.023$)  by turning to a particular spin $1$ Ising 
model designed to have small corrections to scaling. 

 For $N\geq 2$, it is even harder to determine 
the  exponent $\alpha(N)$  in MC simulations, because of the  ambiguity 
in the  separation of the non-divergent singular part of the 
specific heat from the regular background, as argued in Ref.\cite{janke}.

We have also reported a few experimental measurements of the specific heat
 exponent \cite{lipa,ramos,bela,malm} available for $N=1,2,3,4$.

In order to show  quantitatively the 
validity of the hyperscaling relation eq.(\ref{gunt}), 
   our direct  estimates of $\alpha(N)$   
 have  been compared  with the  quantity $2-3 \nu(N)$ also reported in 
 Table I and 
 computed either from  
our extended HT expansions of the 
correlation length\cite{bc3d} for the sc and the bcc lattices,
 or from  the estimates of $\nu(N)$  obtained in  the 
RG approach  by fifth order $\epsilon$-expansion and by seventh order 
FD perturbation expansion\cite{gz2,mur}.
 In conclusion, the hyperscaling relation  
$d\nu(N)=2-\alpha(N)$   appears to 
be reasonably well verified within the 
uncertainties of the data.

\subsection{ Estimates of $R^+_{\xi}(N)$}
We have computed the hyperuniversal combination of critical amplitudes 
$R^+_{\xi}(N)$  by two methods. In 
the first procedure, we evaluate 
the  HT expansion of the quantity

\begin{eqnarray}
 F(N,\beta)=4gq\nu(N)^3\Big(\beta_c(N)\Big)^{9/2} 
\Big(\frac{\xi^2} {\beta}\Big)^{9/2}
 \Big(\frac{d\xi^2}{d\beta}\Big)^{-3} \frac {d^2G(N,\beta)}{d\beta^2}=\\
R^+_{\xi}(N)^3\tau(N)^{2-\alpha(N)-3\nu(N)}(1+O(\tau(N)^{\theta(N)}))
\nonumber \end{eqnarray} 
 at the critical temperature. This computation also provides a good test 
 of hyperscaling: indeed  $F(N,\beta_c(N))=R^+_{\xi}(N)^3$,
 if eq.(\ref{gunt}) holds.
 Here we have found convenient to  use the 
 "simplified" first order  DA's, 
  biased with $\beta^{\#}_c(N)$ and 
 $\theta(N)$,  as described in Ref.\cite{bcgstar},
 and have taken the 
estimates of $\theta(N)$ from Ref.\cite{gz2}.
 We have  reported in Table II only 
the results obtained by this method which is very stable and 
 seems to be  fairly accurate.
 In this case,  our error estimates have to  allow only 
for  the spread of the approximants as well as 
for the uncertainties of 
 $\beta_c(N)$, $\nu(N)$ and  $\theta(N)$. 
The errors quoted mainly derive  from the uncertainties in 
$\theta(N)$, assumed to be generally of the order of $10\%$ and from the 
uncertainties of  $\nu(N)$.
The estimates of $R^+_{\xi}(N)$ obtained by PA's of  $F(N,\beta_c(N))$ are 
systematically smaller by $\approx 5\%$, 
indicating, in our opinion, that the "simplified" DA's 
 are likely to allow more accurately for the sizable negative 
amplitude  corrections to scaling. 
 The usual first order 
DA's biased with $\beta_c(N)$ also seem to lead to less accurate
estimates.

 In the second approach, we obtain $R^+_{\xi}(N)$ from eq.(\ref{egunt}),
after computing separately 
 $A_C(N)$ and $A_{\xi}(N)$  from  the 
specific heat and the correlation length series respectively,
 by  DA's  biased with  
the critical temperatures and exponents.
This second method  leads to results systematically 
smaller (by $\approx 1-2\%$), than those reported in  Table II and  it is 
subject to a  larger uncertainty, 
 due to the necessity of biasing the direct computation of $A_C(N)$  
 also with the exponents $\alpha(N)$, whose relative error
 may be considerable. 

In the same Table  we have also reported 
  the values of  $R^+_{\xi}(N)$   computed 
 via RG\cite{aha} either to second order 
in the $\epsilon$-expansion\cite{hohe} 
 or to fifth  order 
in the FD perturbation expansion\cite{bg}.
 We have also included    earlier estimates obtained 
 in Refs.\cite{stau,fmw,liu1} from the  analysis of  shorter 
 HT series, by the second above mentioned method. 

A recent MC simulation\cite{hp} 
of the Ising model on the sc lattice has determined the universal quantity 
$f^{sc}_s(1) \Big(A^{sc}_{\xi}(1)\Big)^3$ which is closely related to 
$R^+_{\xi}(1)$. Here $f^{sc}_s(1)$
 denotes the amplitude of the singular part of the free energy. 
For convenience, we have translated this result 
  into the estimate of $R^+_{\xi}(1)$ 
reported in Table II, by using the value $\alpha(1)=0.1076(30)$, 
 obtained in the same Ref.\cite{hp} from the hyperscaling eq.(\ref{gunt}).

The values from
 the approximate formula of Bervillier and Godr\`eche have
 been obtained assuming for $\nu(N)$ the FD perturbative results
 of  Ref.\cite{gz2}. We are unable to give sensible 
error estimates in this case,
but it interesting to  quote at least the uncertainties  deriving
 from those of  $\nu(N)$.

 Finally, we should mention that, 
to our knowledge, no other evaluations of $R^+_{\xi}(N)$ for $N=0$ and 
$N=4$ are quoted in the literature.

\subsection{ Estimates of non-universal critical parameters}
 In  Table III,  we have 
reported our 
 estimates of  some non-universal critical parameters, 
for   various values of $N$.
 The inverse critical 
 temperatures $\beta^{\#}_c(N)$,  
 which have been always 
used in the biased analyses of this paper
  were determined in Ref.\cite{bc3d} or taken 
from Refs.\cite{blh,nr90}. 

 The critical amplitudes $A^{\#}_{\xi}(N)$
of the second-moment correlation length 
 were determined in Ref.\cite{bcgstar}. 

 The critical specific energies $U^{\#}(N,\beta_c)$
 and  the critical values of the regular part of the 
specific heat $C^{\#}_{reg}(N,\beta_c(N))$ have been 
obtained by first order DA's biased 
 with $\beta_c(N)$.
 Also  these data are  compatible with the 
 earlier estimates\cite{hp,liu1}.

We have  computed the critical 
amplitudes of the specific heat $A^{\#}_{C}(N)$ in two ways: either  
  indirectly, namely from our estimates of  
$R^+_{\xi}(N)$  by using the knowledge of $A^{\#}_{\xi}(N)$  
and of $\alpha(N)$, or directly, from the specific heat by DA's
biased  with $\beta_c(N)$ and $\alpha(N)$. 
The two methods yield compatible results. We have chosen to report in Table 
III the results of the first approach.  Therefore the 
 relatively large errors  of $A^{\#}_{C}(N)$ 
mainly reflect  the uncertainty 
of  $\alpha(N)$, which, for $N=2$, is so considerable that
 it is not useful to report any estimates in this case. 
(For the same reason we have not
 reported  estimates of  $C^{\#}_{reg}(2,\beta_c(2))$.)
 On the other hand, the uncertainties
 of the products  $\alpha(N)A^{\#}_{C}(N)$
 are more modest and therefore it can be of some interest to quote our 
estimates for $N=2$, namely 
  $\alpha(2)A^{sc}_{C}(2)=0.42(1) $ and 
  $\alpha(2)A^{bcc}_{C}(2)=0.44(1) $. 

We should stress  that here the  meaning of  the errors for 
 $R^+_{\xi}(N)$,  $A^{\#}_{C}(N)$  etc. is not the same as in 
earlier studies, where the errors 
 describe the spread of the estimates in  computations performed at 
 sharply fixed values of $\alpha(N)$ and $\beta_c(N)$.
If, in those computations, we allowed also 
for the uncertainty of  $\alpha(N)$, then  the 
estimates and the errors of $R^+_{\xi}(N)$,  $A^{\#}_{C}(N)$  etc.
  would become completely compatible with our results.    
Therefore, for instance,  we have reported 
in Table III the central values of 
the estimates of $A^{\#}_{C}(1)$  from Ref.\cite{liu1}, 
based on the assignments 
$\alpha(1)=0.104$, $\beta^{sc}_c(1)=0.221630$ 
and $\beta^{bcc}_c(1)=0.157368$, but we have taken the liberty 
of suggesting much larger errors,  which  correspond 
to an indicative $5\%$ uncertainty of $\alpha(1)$.

 Finally, it is  interesting to notice that 
the product $\alpha(N)A^{\#}_{C}(N)$, 
 which is derived  with good accuracy from $R^+_{\xi}(N)$, remains positive 
in the range of $N$ examined here. Therefore, when $\alpha(N)$ 
 changes its sign for $N=\bar N\lesssim 2 $, the same must happen for 
 $A^{\#}_{C}(N)$.
 Analogously  $C^{\#}_{reg}(N,\beta_c)$, which is negative for $N=0,1$,
 has to change sign for $N\geq \bar N$, in order that the maximum of the 
specific heat stays positive.

\section{ Conclusions} 
 We have   analysed  our  extended  HT expansion of $G(N,\beta)$
 for the sc and bcc lattices in order to   update the  direct 
 estimates 
of the critical exponent $\alpha(N)$ 
and of the hyperuniversal combination of amplitudes $R^+_{\xi}(N)$, 
over a  range of values of $N$.  

 Due to the smallness of $\alpha(N)$ and to the limited effective length  
of the series, the relative accuracy of 
 our extrapolations is still generally inferior to that already 
achieved in our
 recent HT studies of the  susceptibility and of the 
 correlation length critical exponents\cite{bc3d}. 
However,  within  the  error limits, 
  the main predictions of universality, hyperuniversality and 
 hyperscaling appear to be well verified and the overall 
agreement between the  HT and the RG  estimates of the 
universal observables  is good.

\acknowledgments 
This work has been partially supported by MURST. 
We thank Prof. J. Zinn-Justin for a very useful correspondence.

\begin{table}
\caption{ In the first six lines we have reported 
 the direct HT estimates of the critical exponents 
$\alpha(N)$  obtained  in this work by various routes:
 by  first order DA's of the specific heat biased with $\beta_c(N)$; 
 by similarly biased  extrapolation 
 of  ratios of the specific heat series coefficients
 and by   second order DA's 
of $d\chi/d\beta$ biased with $\beta_c(N)$ and $\beta^{af}_c(N)$. 
We have then reported 
earlier direct  estimates from shorter HT series, 
 some direct MC determinations,
 and a few  experimental measures.
 For each value of $N$, our estimates of $\alpha(N)$ 
 have to be compared with the quantity  2-3$\nu(N)$ 
 reported in the last four lines and obtained either from  our previous 
HT study of the   correlation 
length series or from  RG estimates 
 via $\epsilon$-expansion and  via FD perturbative expansion.} 
\label{tabella1}
\begin{tabular}{llllll}
 $N$     & 0      & 1   &   2 &  3   &  4\\
\tableline
$C^{sc}_H(N,\beta)$ DA&0.24(1) &0.103(8) &$-0.014(9)$ 
&$-0.11(2)$ &$-0.22(4)$\\
$C^{bcc}_H(N,\beta)$ DA&0.23(1) &0.105(9) &$-0.019(8)$
&$-0.13(2)$&$-0.25(3)$  \\
$C^{sc}_H(N,\beta) $Ratio Ext.&0.236(8)&0.104(6)&$-0.020(8)$
&$-0.15(^{+4}_{-1})$&$-0.27(^{+6}_{-1})$\\
$C^{bcc}_H(N,\beta) $Ratio Ext.&0.233(8)&0.106(6)&$-0.022(6)$
&$-0.16(^{+3}_{-1})$&$-0.29(^{+6}_{-1})$\\
$\frac {d\chi^{sc}} {d\beta}(N,\beta)$ DA    &0.239(8)
&0.13(3)&0.02(3)&$-0.13(3)$&$-0.24(3)$                \\
$\frac {d\chi^{bcc}} {d\beta}(N,\beta)$ DA   &0.233(6)&0.107(8)
&$-0.01(2)$&$-0.138(8)$&$-0.23(2)$  \\
  $\chi^{bcc}(1,\beta)$ DA Ref.\cite{fich} & &0.105(7) & & &   \\
  $\chi^{bcc}(1,\beta)$ DA Ref.\cite{nr90} & &0.105(7) & & &   \\
 $C_H(N,\beta)$ DA Ref.\cite{chfi} & &0.125(25) &$-0.02(3)$ &$-0.22(4)$& \\
MC Refs.\cite{def,blh,janke}
&0.275(15)&0.125(23)&&$-0.23(16)$& \\   
 Exper. Refs.\cite{gz2,blh,lipa,ramos,bela,malm}& &0.107-0.112
&$-0.01285(38)$&$-0.135(2)$ &$-0.20(5)$  \\
  2-3$\nu(N)$ DA sc Ref.\cite{bc3d}&0.2366(18)
&0.1055(24) &$-0.025(6)$&$-0.148(6)$&$-0.277(9)$ \\
 2-3$\nu(N)$ DA bcc Ref.\cite{bc3d}&0.2363(18)
&0.1076(15)&$-0.022(6)$&$-0.142(6)$&$-0.268(9)$\\
 2-3$\nu(N)$  $\epsilon$-expans. Ref.\cite{gz2}&0.2375(54)&0.1121(78)
&$-0.0055(120)$&$-0.115(16)$&$-0.211(24)$\\ 
 2-3$\nu(N)$  FD-pert. Ref.\cite{mur}&0.235(3)&0.109(4)
&$-0.011(4)$ &$-0.122(10)$&$-0.223(18)$\\ 
\end{tabular}
\end{table}

\begin{table}
\caption{ Estimates of the hyperuniversal quantity $R^+_{\xi}(N)$. 
The results of 
 our HT series computation are compared with  RG estimates via 
$\epsilon$-expansion or via fixed-dimension 
perturbative expansion, with a
 heuristic approximate formula  and with experimental measures.} 
\label{tabella2}
\begin{tabular}{llllll}
 $N$   & 0      & 1   &   2 &  3   &  4\\
\tableline
HT sc (this work)&0.258(3)&0.273(4)&0.361(4)&0.431(5)&0.497(6)\\
HT bcc (this work)&0.258(3)&0.272(4)&0.362(4)&0.433(5)&0.500(6)\\
RG $\epsilon$-expans. Ref.\cite{ber}& &0.27 &0.36 &0.42&    \\
RG FD-pert. Refs.\cite{aha,bg} & & 0.270(1) &0.361(2)&0.435(2) & \\
HT  Refs.\cite{liu1,fmw,stau} & &0.2659(7)& 0.36(1)&0.42 & \\
 MC Ref.\cite{hp}& &0.2685(60)& & & \\ 
$\nu(N)\big(\frac {N} {4\pi}\big)^{1/3}$ Ref.\cite{bg}
&&0.271(1)&0.363(1)&0.439(2) &0.506(4) \\
 Exper. Refs.\cite{stau,bg,kumar} & &0.25-0.32 & &0.40-0.45 &  \\
\end{tabular}
\end{table}

\begin{table}
\caption{ Estimates of non-universal parameters.
 We report the critical inverse temperatures 
$\beta^{\#}_c(N)$ always used in our biased 
procedures, the critical amplitudes 
$A^{\#}_{\xi}(N)$ and $A^{\#}_{C}(N)$, the  critical 
 specific energies $U^{\#}(N,\beta_c)$ and the critical values of the 
regular part of $C^{\#}_H(N,\beta)$.} 
\label{tabella3}
\begin{tabular}{llllll}
 $N$     & 0      & 1   &   2 &  3   &  4\\
\tableline
$\beta^{sc}_c(N)$ HT  Refs.\cite{bc3d,blh}
&0.213493(3)&0.2216544(3)&0.45419(3)&0.69305(4)&0.93600(4)\\
$\beta^{bcc}_c(N)$ HT  Refs.\cite{bc3d,nr90}&0.153128(3)
&0.157373(2)&0.320427(3)&0.486820(4)&0.65542(3)\\
$A^{sc}_{\xi}(N)$ HT
Ref.\cite{bcgstar}&0.5101(3)&0.5027(3)&0.4814(3)&0.4541(3)&0.4155(3)\\ 
$A^{bcc}_{\xi}(N)$ HT Ref.\cite{bcgstar}
 &0.4846(2)&0.4659(2)&0.4371(2)&0.4072(2)&0.3691(2) \\ 
$A^{sc}_{C}(N)$ (this work)  &0.546(8)&1.49(5)&   &-6.0(6)&-6.5(3)\\ 
$A^{bcc}_{C}(N)$  (this work) &0.481(6)&1.43(4)&       &-6.5(6)&-7.2(3)\\ 
$A^{sc}_{C}(N)$ MC Ref.\cite{hp}& &1.45(9)&   &&\\ 
$A^{sc}_{C}(N)$ HT Ref.\cite{liu1}  &    &1.464(90)&       &   &      \\ 
$A^{bcc}_{C}(N)$ HT  Ref.\cite{liu1} &    &1.431(80)&       &   &     \\ 
$U^{sc}(N,\beta_c)$  (this work)&-1.004(3)&-0.991(1)&-0.990(3)
&-0.991(3) &-0.994(4) \\
$U^{bcc}(N,\beta_c)$ (this work)&-1.0990(2)&-1.0903(6)&-1.0896(8)
&-1.0919(4)&-1.0951(2)\\
$U^{sc}(N,\beta_c)$ HT  Ref.\cite{liu1} &  &-0.9902(1)&  &   &  \\
$U^{sc}(N,\beta_c)$ MC  Ref.\cite{blh} &  &-0.9904(8)&  &   &  \\
$U^{bcc}(N,\beta_c)$ HT  Ref.\cite{liu1}&  &-1.0904(1)&   &   &  \\
$C^{reg}_H(N,\beta_c)$ sc  (this work)&-0.66(3)&-1.67(3)&
  &4.9(4)&4.2(3)  \\
$C^{reg}_H(N,\beta_c)$ MC sc Refs.\cite{hp,janke}&  
&-1.64(11)& &5.79(12) &\\
$C^{reg}_H(N,\beta_c)$ MC sc Ref.\cite{cfl}&  && &5.70(12) &\\
$C^{reg}_H(N,\beta_c)$ bcc (this work)
&-0.68(2)&-1.64(3)&  &5.2(4)&4.3(3) \\
$C^{reg}_H(N,\beta_c)$ MC bcc Ref.\cite{cfl}&  && &5.54(14) &\\
\end{tabular}
\end{table}

\appendix

\section{The nearest neighbor correlation function on the sc lattice}

The HT expansion coefficients of the 
nearest neighbor correlation function on the sc lattice are

\scriptsize

\[a_1(N)= \frac{1} {N} \]

\[a_3(N)= \frac{8 + 3N} {N^3(2 + N)} \]

\[a_5(N)= \frac{352 + 168N + 22N^2} {N^5(2 + N)(4 + N)}\]

\[a_7(N)= \frac{105984 + 154752N + 85056N^2 + 21960N^3 + 2754N^4 + 135N^5}
	{N^7(2 + N)^3(4 + N)(6 + N)}\]

\normalsize
For the coefficients which follow, it is typographically
more convenient to set $a_r(N)=P_r(N)/Q_r(N)$ and to
 tabulate separately the numerator polynomial $P_r(N)$
 and the denominator polynomial $Q_r(N)$,
\scriptsize
  
\[P_9(N)=12349440 + 17871360N + 10010240N^2 + 2751680N^3 + 405776N^4 +
     30876N^5 + 954N^6\]

\[Q_9(N)=N^9(2 + N)^3(4 + N)(6 + N)(8 + N)\]

\[P_{11}(N)=124861808640 + 364560318464N + 467027804160N^2 
+ 345395589120N^3 + 163465120768N^4 + 51937662976N^5  
+ \]\[ 11315941120N^6 +
 1694683328N^7 + 171418048N^8 + 11171800N^9 + 422520N^{10} + 7026N^{11}\]

\[Q_{11}(N)=N^{11}(2 + N)^5(4 + N)^3(6 + N)(8 + N)(10 + N)\]

\[P_{13}(N)=24917940633600 + 71794651299840N + 91099400634368N^2 +
     67066306363392N^3 + 31821500096512N^4 +\]\[ 10242128590848N^5  +
     2295320471552N^6 + 361789563776N^7 + 39924856512N^8 +
     3014946464N^9 +\]\[ 148081312N^{10} + 4249712N^{11} + 53892N^{12}\]

\[Q_{13}(N)=N^{13}(2 + N)^5(4 + N)^3(6 + N)(8 + N)(10 + N)(12 + N)\]

\[P_{15}(N)=867654721119191040 
 + 3616829986427633664N + 6891583739428601856N^2 +
     7957383254837821440N^3 +\]\[ 6225913571872604160N^4 +
     3498334649912000512N^5 + 1460523056889888768N^6 +\]\[
     462563223592566784N^7 + 112521154820349952N^8 +
     21154253531684864N^9+\]\[  3076240360587264N^{10} 
+ 344376491174400N^{11} +
     29339259414560N^{12} + 1863409665456N^{13}+\]\[  85223778256N^{14} +
     2644451768N^{15} + 49679114N^{16} + 425007N^{17}\]

\[Q_{15}(N)=
N^{15}(2 + N)^7(4 + N)^3(6 + N)^3(8 + N)(10 + N)(12 + N)(14 + N)\]

\[P_{17}(N)= 3948322260048528015360 + 18226598259687325433856N  +
     38988021723789936033792N^2 +\]\[ 51323869690127645147136N^3 +
     46583550742458833829888N^4 + 30960681462370651865088N^5 +\]\[ 
     15623251635335279411200N^6 + 6126114771192359944192N^7 +
     1895134340075627937792N^8+\]\[  467022808981231222784N^9 +
     92186181864442351616N^{10} + 14603683596490825728N^{11}+\]\[ 
     1853863098715137024N^{12} + 187606064202660864N^{13} +
     14988669525495552N^{14}+\]\[  930810012214464N^{15} 
   + 43862323328864N^{16} +
     1510537882592N^{17} +\]\[ 35726075472N^{18} + 516586876N^{19}
  + 3426610N^{20}\]

\[Q_{17}(N)= 
N^{17}(2 + N)^7(4 + N)^5(6 + N)^3(8 + N)(10 + N)(12 + N)(14 + N)
     (16 + N)\]

\[P_{19}(N)= 330768394077031316324352000 +
   1921489492806829461838233600N +\]\[  5244352748560893054120099840N^2 +
   8943585498141047607892377600N^3+\]\[  10692857932684576404138885120N^4 +
   9533746112508667703922262016N^5+ \]\[  6584053425730588600199806976N^6 +
   3611602580377927390173593600N^7 +\]\[  1601207739146215800698830848N^8 +
   580892231405628018430836736N^9+ \]\[  173958231237568098749120512N^{10} +
   43263481821264025859260416N^{11}+\]\[  8970159797560936461959168N^{12} +
   1553563398790168428314624N^{13}+ \]\[  224789497420511579963392N^{14} +
   27127717091438526734336N^{15}+ \]\[  2720424198488158732288N^{16} +
   225326679276573418496N^{17} + 15276530595902585344N^{18} +\]\[
   836884471109722496N^{19} + 36371584704297344N^{20} +
   1221231831603552N^{21} +\]\[ 30441535564576N^{22} + 528211335752N^{23} +
   5666057752N^{24} + 28113366N^{25}\]

\[Q_{19}(N)= 
N^{19}(2 + N)^9(4 + N)^5(6 + N)^3(8 + N)^3(10 + N)(12 + N)(14 + N)
     (16 + N)(18 + N)\]

\[P_{21}(N)=122469323387965953278371430400 +
     703968135713996874968318607360N+ \]\[ 
     1902018774401482372925381148672N^2 +
     3212833143009475621586001199104N^3+ \]\[ 
     3807421545781558189263566143488N^4 +
     3367773291711700412520831385600N^5+ \]\[ 
     2309856527527444083001140445184N^6 +
     1260025811629511034804556005376N^7+ \]\[ 
     556434796744868231529151594496N^8 +
     201462619317391517373909958656N^9+\]\[ 
     60352333626138484849753718784N^{10} +
     15057212981368189621789261824N^{11}+\]\[ 
 3142458232040362531962355712N^{12} 
+ 550072000489168244280950784N^{13}+ \]\[
 80841983595825376237305856N^{14} + 9969253742422581036474368N^{15}+\]\[ 
 1029177452713566097747968N^{16} + 88565456715317390606336N^{17}+\]\[ 
 6311161666273241710592N^{18} + 368858480778492513280N^{19}+\]\[ 
 17443565419605911296N^{20} + 654851921510017152N^{21}+ 
     18987446839217536N^{22}\]\[ + 408099174234848N^{23}+
     6085840871680N^{24} +
     55773063792N^{25} + 233966556N^{26}\]

\[Q_{21}(N)=N^{21}(2 + N)^9(4 + N)^5(6 + N)^3(8 + N)^3(10 + N)(12 + N)
(14 + N)(16 + N)(18 + N)(20 + N)\]

\normalsize

	In particular, for $N=0$ [the  SAW model],
	we have (in terms of the variable 
	$\tilde \beta=\beta/N$)

\scriptsize
\[G(0,\tilde\beta)= 
                                \tilde\beta 
                              + 4 \tilde\beta^{ 3} 
                              +44 \tilde\beta^{ 5} 
                           +552 \tilde\beta^{7} 
                           +8040 \tilde\beta^{ 9} 
                          +127016 \tilde\beta^{ 11} 
			 +2112320 \tilde\beta^{ 13}
			+36484128\tilde\beta^{ 15}
			+648529392\tilde\beta^{ 17}\]\[
			+ 11790401800\tilde\beta^{ 19}
    +218273957968\tilde\beta^{ 21}+...\]

\normalsize

	For $N=1$ [ the spin 1/2 Ising model], we have 

\scriptsize
\[G(1,\beta)= 
\beta+ 11/3\beta^{3}+ 542/15\beta^{5}+ 123547/315\beta^{7}+
 14473442/2835\beta^{9}+ 
  11336607022/155925\beta^{11}+  605337636044/552825\beta^{13}+\]\[
 10976336338579019/638512875\beta^{15}+ 
 3022947654230404442/10854718875\beta^{17}+ 
  8582760723898537620322/1856156927625\beta^{19}+  \]\[
  15262009695163033631128084/194896477400625\beta^{21}+...   \]

\normalsize

	For $N=2$ [ the  XY model], we have 

\scriptsize
\[G(2,\beta)= 
1/2\beta+ 7/16\beta^{3}+ 97/96\beta^{5}+ 5103/2048\beta^{7}+
 459719/61440\beta^{9}+ 218788559/8847360\beta^{11}+  
 3579816967/41287680\beta^{13}+\]\[ 20154248931151/63417876480\beta^{15}+
  4126827327908711/3424565329920\beta^{17}+ 
2142771095208749011/456608710656000\beta^{19}+  \]\[
  562665453010146198199/30136174903296000\beta^{21}+... \]

\normalsize

	For $N=3$ [the classical Heisenberg model], we have

\scriptsize
\[G(3,\beta)= 
1/3\beta+ 17/135\beta^{3}+ 1054/8505\beta^{5}+
 80909/637875\beta^{7}+ 95738/601425\beta^{9}+ 
  5992817408726/27152760009375\beta^{11}+\]\[ 
 11357358327572/34910691440625\beta^{13}+ 
  156550175755271443/311577921107578125\beta^{15}+ \]\[
  190956190202826883834/237337400087308828125\beta^{17}+  
  56535690823720347706912558/42645970734688086781640625\beta^{19}+ \]\[
  358752594209204675460504716/160503926219644253887265625\beta^{21}+... \]

\normalsize

\section{The nearest neighbor correlation function on the bcc lattice}

\normalsize
The HT expansion coefficients of the 
nearest neighbor correlation function on the bcc lattice are

\scriptsize

\[a_1(N)= \frac{1} {N} \]

\[a_3(N)= \frac{24 + 11N}{N^3(2 + N)}\]

\[a_5(N)= \frac{1776 + 1044N + 152N^2}{N^5(2 + N)(4 + N)}\]

\[a_7(N)= \frac{ 1050624 + 1713024N + 1062432N^2 + 312600N^3 
+ 44090N^4 + 2395N^5}
   {N^7(2 + N)^3(4 + N)(6 + N)}\]

\normalsize
For the coefficients which follow, it is typographically
more convenient to set $a_r(N)=P_r(N)/Q_r(N)$ and to
 tabulate separately the numerator polynomial $P_r(N)$
 and the denominator polynomial $Q_r(N)$,
\scriptsize

\[P_9(N)= 237680640 + 391630080N + 251136960N^2 
+ 79995360N^3 + 13572456N^4 +
     1175956N^5 + 40904N^6\]

\[Q_9(N)= N^9(2 + N)^3(4 + N)(6 + N)(8 + N)\]

\[P_{11}(N)= 4657615994880 + 14662439436288N + 20306810757120N^2 +
     16297064577024N^3 + 8408736450048N^4 +\]\[ 2927305709568N^5 +
     701958299776N^6 + 116098602304N^7 + 13001482080N^8 +
     940546304N^9 + 39618896N^{10} + 737112N^{11}\]

\[Q_{11}(N)=N^{11}(2 + N)^5(4 + N)^3(6 + N)(8 + N)(10 + N)\]

\[P_{13}(N)= 1804392176025600 + 5660420904714240N + 7838369893122048N^2 +
     6320116029308928N^3 + 3299174287417344N^4 + \]\[1173899872406016N^5 +
     292101988094976N^6 + 51294669578688N^7 + 6322737698272N^8 +\]\[
  534749498288N^9 + 29520640808N^{10} + 956957440N^{11} + 13799232N^{12}\]

\[Q_{13}(N)=N^{13}(2 + N)^5(4 + N)^3(6 + N)(8 + N)(10 + N)(12 + N)\]

\[P_{15}(N)= 122002510248369192960 + 543062014542747795456N +
     1106788272284626845696N^2 +\]\[ 1369759313298192334848N^3 +
     1151523649799700086784N^4 + 697153536634263011328N^5 +\]\[
     314532460294909476864N^6 + 107981066371807617024N^7 +
     28558819799096193024N^8 +\]\[ 5854384426156062720N^9 +
     930798833517987840N^{10} + 114224602657806848N^{11} +\]\[
 10696888031248800N^{12} 
+ 749147616393328N^{13} + 37927213609168N^{14} +\]\[
     1309142853624N^{15} + 27530444114N^{16} + 265776699N^{17}\]

\[Q_{15}(N)=
N^{15}(2 + N)^7(4 + N)^3(6 + N)^3(8 + N)(10 + N)(12 + N)(14 + N)\]

\[P_{17}(N)= 1078176657766748635791360 + 5311087820388289065517056N +\]\[
     12134778449671899486093312N^2 + 17084121119131816471560192N^3 +\]\[
     16608988472236389552881664N^4 + 11844719796924403830226944N^5 +\]\[
     6426005983019854016544768N^6 + 2714739161543420662382592N^7 +\]\[
     906815315440850230886400N^8 + 241845902161952176398336N^9 +\]\[
     51782172208072962975744N^{10} + 8918298897939850518528N^{11} +\]\[
     1233687735206166823424N^{12} + 136374954925264681472N^{13} +\]\[
     11933805821587623936N^{14} + 814292868353822272N^{15} +
     42326351689562720N^{16} +\]\[ 1615854325367776N^{17} + 
	42636630380712N^{18} +
     693565371332N^{19} + 5232689960N^{20}\]

\[Q_{17}(N)=N^{17}(2 + N)^7(4 + N)^5(6 + N)^3(8 + N)(10 + N)(12 + N)(14 + N)
     (16 + N)\]

\[P_{19}(N)= 175433838338452762972599091200 +
     1078203808265217149807540305920N +\]\[
     3116173607652516436321480212480N^2 +
     5633274990887456528054857236480N^3 +\]\[
     7147764845163022335548483174400N^4 +
     6772123960394212475971911548928N^5 +\]\[
     4976697709965417132847811002368N^6 +
     2909196906278057822968464015360N^7 +\]\[
     1376618167955575605597824876544N^8 +
     533877242182341657037822230528N^9 +\]\[
     171186415133933732406244147200N^{10} +
     45658929843087077741062520832N^{11} +\]\[
     10169112743416038967608705024N^{12} +
     1894893729303875646595006464N^{13} 
+\]\[ 295461155711988902515834880N^{14} +
     38487351224017884772339712N^{15} 
+\]\[ 4173283404716049615550464N^{16} +
     374483776338165994216448N^{17} +\]\[ 27568453015797659725824N^{18} +
     1644491785569570191872N^{19} +\]\[ 78098433373280888576N^{20} +
     2878619676002249280N^{21} +\]\[ 79247785021379008N^{22} +
     1531012629840624N^{23} + 18487394459632N^{24} + 104841714952N^{25}\]

\[Q_{19}(N)=N^{19}(2 + N)^9(4 + N)^5(6 + N)^3(8 + N)^3
 (10 + N)(12 + N)(14 + N)(16 + N)(18 + N)\]

\[P_{21}(N)=  (126175500888039348819208018329600 +
     771182011973845295293568951255040N +\]\[
     2217488378116952516842355989413888N^2 +
     3990433665646539615566917678399488N^3 +\]\[
     5043598010921343492445826797535232N^4 +
     4763919940749294277936547050291200N^5 +\]\[
     3493715661682982610876998803783680N^6 +
     2040596945264227996384033763229696N^7 +\]\[
     966223362153625194801269330411520N^8 +
     375629302981996871057940890517504N^9 +\]\[
     120996637012851297773150580768768N^{10} +
     32504072855072819274267570733056N^{11} +\]\[
     7314027149319445947024024403968N^{12} +
     1382157562600983080851995279360N^{13} +\]\[
     219565426180384061435124916224N^{14} +
     29302944828757838601074274304N^{15} +\]\[
  3278043851101943224116459520N^{16} 
+ 306114536101034343439527936N^{17} +\]\[
  23712149887809097730532352N^{18} + 1509817181912122244980224N^{19} +\]\[
  78017870011349442092288N^{20} + 3213483265751133834688N^{21} +
  102816420657321623712N^{22} +\]\[ 2458334158401005552N^{23} +
  41257373964220632N^{24} + 432739125346952N^{25} + 2130772922816N^{26}\]

\[Q_{21}(N)= 
N^{21}(2 + N)^9(4 + N)^5(6 + N)^3(8 + N)^3(10 + N)(12 + N)(14 + N)
     (16 + N)(18 + N)(20 + N)\]

\normalsize

	In particular, for $N=0$ [the  SAW model],
	we have (in terms of the variable 
	$\tilde \beta=\beta/N$)

\scriptsize
\[G(0,\tilde\beta)= 
                                \tilde\beta 
                              +12 \tilde\beta^{ 3} 
                              +222 \tilde\beta^{ 5} 
                           +5472 \tilde\beta^{7} 
                           + 154740 \tilde\beta^{ 9} 
                          +4737972 \tilde\beta^{ 11} 
			 + 152960220 \tilde\beta^{ 13}
			+  5130099672\tilde\beta^{ 15}
			+ 177095284092\tilde\beta^{ 17}\]\[
			+ 6253425298080\tilde\beta^{ 19}
     + 224879383796232\tilde\beta^{ 21} +...\]

\normalsize

	For $N=1$ [ the spin 1/2 Ising model], we have 

\scriptsize
\[G(1,\beta)=
\beta+ 35/3\beta^{3}+ 2972/15\beta^{5}+ 279011/63\beta^{7}+
 46439636/405\beta^{9}+ 
  100877055128/31185\beta^{11}+ 587703506650264/6081075\beta^{13}+\]\[ 
  10981652882712713/3648645\beta^{15}+ 
1049923978894758374012/10854718875\beta^{17}+ 
  1182698210781462071363672/371231385525\beta^{19}+ \]\[
  2980059927747623321534851312/27842353914375\beta^{21}+...\]

\normalsize

	For $N=2$ [ the  XY model], we have 

\scriptsize
\[G(2,\beta)=
1/2\beta+ 23/16\beta^{3}+ 559/96\beta^{5}+ 187645/6144\beta^{7}+
 11417419/61440\beta^{9}+ 
  10934199853/8847360\beta^{11}+\]\[ 1081218105839/123863040\beta^{13}+ 
  4085878131871327/63417876480\beta^{15}+
 1683908448367350071/3424565329920\beta^{17}+ \]\[
  753925204192677068291/195689447424000\beta^{19}+ 
  929152049503798552678997/30136174903296000\beta^{21}+...\]

\normalsize

	For $N=3$ [ the classical Heisenberg model], we have 

\scriptsize
\[G(3,\beta)=
1/3\beta+ 19/45\beta^{3}+ 2092/2835\beta^{5}+
 349939/212625\beta^{7}+ 2147444/505197\beta^{9}+ 
  108732988464808/9050920003125\beta^{11}+\]\[
 9339742669288/258597714375\beta^{13}+ 
  35412600932786885263/311577921107578125\beta^{15}+\]\[ 
  35816645375345371477924/96693014850385078125\beta^{17}+ 
 5880568448944900843943527784/4738441192743120753515625\beta^{19}+ \]\[
1137371495914136837811604445344/267506543699407089812109375\beta^{21}+...\]

\end{document}